# What Can Next Generation Ultrafast Light Sources Tell Us About Quantum Materials?

**Allan S. Johnson***

*IMDEA Nanoscience, Calle Faraday 9, 28049 Madrid, Spain. ORCID 0000-0002-0711-708X*

## Summary

Ultrafast spectroscopy of quantum materials has emerged as a principle application of modern ultrafast optical sources. Measurements made with ultrashort pulses allow the normally strongly coupled internal degrees of freedom inside these systems to be de-coupled in the time-domain, revealing the key interactions for the different orders that lead to dramatic emergent phenomena. This has naturally motivated a wide range of experiments, and quantum materials have now been studied with ultrafast sources spanning from the hard X-ray to the THz, with fields intense enough to create non-equilibrium states, and using pulses down to the attosecond duration. As the optical spectroscopist's toolbox becomes increasingly complete, it is worth pausing and reflecting on what are the capabilities that we currently lack for examining quantum materials, and what are the problems that require the next generation of ultrafast sources now coming on-line or being designed. Here I will offer an overview of the remaining missing capabilities in ultrafast optical sources as required for studying quantum materials, and identify some outstanding problems where I believe these new sources can provide decisive impact in our understanding of quantum and strongly-correlated materials.

## Main Text

In the last two decades ultrafast spectroscopy has expanded from the study of relatively well understood systems like band-semiconductors or metals to the more mysterious world of quantum materials (1,2). These materials exhibit properties and emergent phenomena arising from quantum effects and collective interactions, which can challenge established theoretical frameworks and may require approaches beyond conventional models for their full understanding. Ultrafast spectroscopy has been applied to practically every class of problem related to quantum materials, including superconductors (3–5), Dirac materials (6–9), Mott-insulators (10–12), Moiré materials (13,14), charge density waves (15,16), and many more, as detailed in existing reviews (3,17–20). Equally exciting is the discovery that ultrafast excitation can drive quantum materials into entirely new states which cannot be reached through thermodynamic pathways; examples include light-induced superconductivity (21,22), photo-induced Chern insulators (23), topologically protected charge-density wave arrangements (24,25), and emergent ferroelectric and ferromagnetic states (26,27).

The great advantage of ultrafast spectroscopy in studying or controlling quantum materials is its ability to separate couplings between different degrees of freedom in the time domain (3,17,28). Normally, inside quantum materials, the different internal degrees of freedom (electrons, spins, lattice, orbitals, etc.) are strongly coupled, meaning that a perturbation to one degree of freedom leads to perturbation to all the others (1). This makes it difficult to ascertain which degrees of freedom are truly important for a given emergent phenomena. In the time domain this flow of energy can be "outrun", and the different couplings and their effects tracked independently. In the case of control experiments, it similarly allows for targeted control over separate degrees of freedom. For example, photo-excitation generally deposits energy into the electrons, and the flow of energy into the lattice and spins can then be tracked (29,30) to report on when emergent phenomena like coupled structural-magnetic transitions occur, or to melt electronic order while leaving the lattice structure untouched (31). Furthermore, following the seminal work of Allen in 1987 (28), the rate of energy transfer itself can be used to ascertain the coupling strengths between the different degrees of freedom provided the subsystems can be assigned well-defined (e.g. quasi-thermal) distributions and identified uniquely from experimental observables; otherwise, fitted decay constants remain phenomenological. This capability is rather unique, and is conceptually behind ideas using ultrafast spectroscopy to look for bottleneck rates in ultrafast phase transitions (32), extract electron-phonon scattering rates and potential barriers (33), and examine superconducting pair formation mechanisms (18). While limited temporal information can be extracted from spectroscopic probes under appropriate assumptions (34), real time measurements are required to recover these more complex effects.

*allan.johnson@imdea.org

Applying ultrafast spectroscopy to quantum materials relies, however, on being able to both excite and track the independent degrees of freedom separately and on faster timescales than the couplings/thermalization. Addressing particular degrees of freedom can sometimes be done through selection of the photon energy as phonons, electronic transitions and other degrees of freedom often appear at distinct spectral windows (35,36); this approach is also beneficial for ultrafast optical control, where accessing emergent states typically relies on selective excitation of particular degrees of freedom (37–39). Notably, because conventional near-infrared excitation deposits energy far above most collective energy scales, this energy reaches the low-energy degrees of freedom only through a cascade of processes. Such experiments nonetheless remain powerful probes of screening, energy redistribution, and bottlenecks, but the connection between the transient state and low-energy equilibrium physics is non-trivial(40) and the specific relationship must be established in each case. Beyond optical measurement, probing the underlying degrees of freedom can also be done through nonlinear effects like second harmonic generation for tracking inversion symmetry breaking (41) or the magnetic-optical Kerr effect for tracking magnetic spin order (42). Perhaps most critically, with the advent of X-ray free electron lasers, ultrafast science can now be combined with a wide range of condensed matter's most powerful tools like X-ray diffraction (17,30), resonant X-ray inelastic scattering (RIXS) (43,44), pair-distribution function analysis (45) and high-resolution X-ray imaging (46,47), allowing momentum and real-space resolved measurements of many types of phenomena. Other sources like ultrafast electron beams (48) are also an important complement to optical methods, but are beyond the scope of this perspective.

Despite these advances, the underlying physics of quantum materials remains mysterious and ultrafast probes have often opened as many questions as they answer by introducing the complications of dynamics. New spectroscopic probes are being continuously developed and applied to the condensed phase (49): notable frontiers include combining optical sources with scanning probes for deeply sub-wavelength imaging (50), light-wave driven spectroscopies (51) and attosecond pump-probe (52). It is worth noting that interpreting these experimental signals, even for conventional probes, is far from trivial. In equilibrium, established relations connect observables to intrinsic quantities, such as reflectivity to optical conductivity or ARPES intensity to the spectral function. Out of equilibrium, however, the pump-prepared state is not time-translation invariant, so the response to even a weak probe is generally a two-time quantity. Reporting it as a transient conductivity, spectral function, or susceptibility therefore assumes the state is approximately stationary over the interaction window with the probe, and the result can depend on the probe envelope, delay convention, and frequency window. In general the signal in such experiments is a nonlinear, protocol-dependent response. More detailed descriptions of how to link these different experimental observables to underlying physical quantities can be found in selected reviews (35,37,49). In general these new probes are each enabled by new types of light sources as much as by methodological advances, and so it is well worth to ask, what is still missing from the ultrafast spectroscopist's toolbox for studying quantum materials, and what problems can new technological developments in these directions be reasonably expected to solve?

Classical coherent light sources can be described within a framework comprising five fundamental degrees of freedom—wavelength, pulse energy, repetition rate, pulse duration, and polarization (Figure 1)—together with the spatial-temporal profiles of the optical field. Numerous other representations of these five parameters are possible (for instance power, frequency, fluence, etc.) but considering only coherent light sources, these five are sufficient to discuss the cases of interest. In considering new light sources for studies of quantum materials we can further group wavelength and pulse energy for a simple reason – we can now routinely generate light at basically any wavelength of interest from the UV to the radio wave. The real question is whether we can do so with enough energy to drive processes of interest for nonlinear spectroscopies or control schemes. For many wavelengths it is relatively simple to provide pulse energies so high they will melt the delicate quantum states of interest, if not completely destroy the material (47), and so when we speak of new wavelength sources what we really mean is new sources with high pulse energy at those wavelengths. This now largely concerns the very far-infrared/THz, where fields typically remain "weak" in comparison to material damage (53), and in the UV where intense sources have only emerged recently (54). While UV pulses are of tremendous interest for studies of molecular and liquid systems (55,56), for quantum materials this spectral range is typically of less interest as it probes transitions far away from the Fermi level where the action happens. UV sources have traditionally been invaluable for photoelectron measurements in solids, but are generally outperformed by XUV or X-ray radiation now available at with tabletop sources (57).

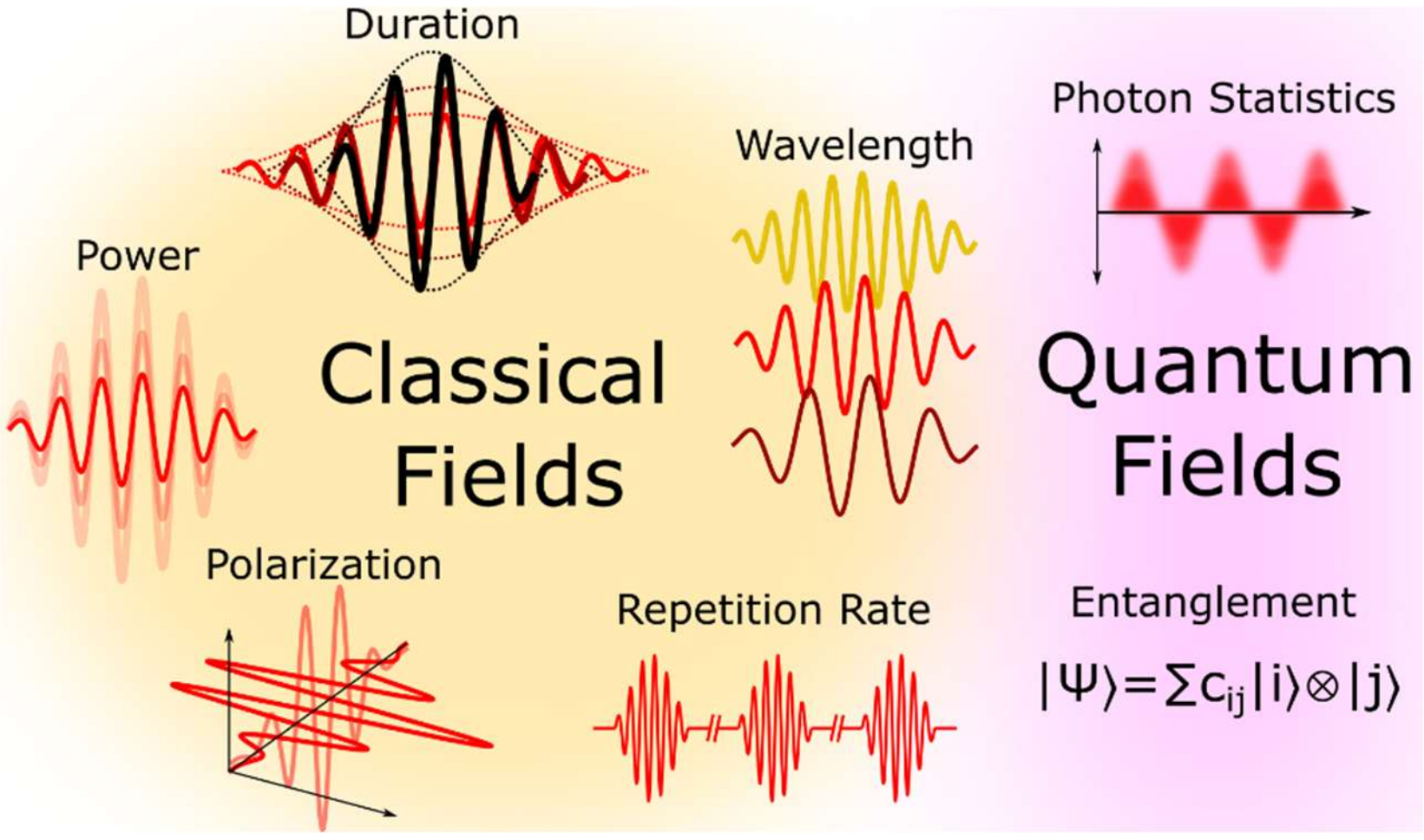

***Figure 1: Properties of classical and quantum ultrafast light sources.*** *Classical sources can be described by their duration, power, wavelength, polarization and repetition rate, as one non-unique representation. Quantum fields are distinguished through their photon statistics and potential non-classical correlations.*

In this perspective I will present how new emerging light sources can impact spectroscopy in quantum materials. In order to maintain a reasonable scope I will not discuss the use of structured light, which may play an important role for the study of quantum materials particularly as it becomes accessible with X-rays and the structuring approaches the fundamental length scales in materials (58). Similarly I will not discuss polarization, though polarization control remains challenging for large bandwidth pulses across basically all regions of the electromagnetic spectrum. With this in mind I will focus on four key cases: the emergence of practical attosecond or single-digit-femtosecond pulses, the scaling of intense femtosecond sources to high repetition rates (100s of kHz to MHz), the use of high-field THz and far-IR sources, and the potential of ultrafast pulses with non-classical properties, namely non-coherent state photon statistics and entanglement (Figure 1). This last topic is conceptually and technologically distinct from the classical properties of light, and I will accordingly examine it separately. In each case I will endeavour to highlight particular key open problems where these new

capabilities – those that have become available only in the past few years or that will emerge in the decade to come- could play an important role.

**Attosecond and Few-Femtosecond Pulses**

There has in the past been reservation in the quantum materials community regarding the importance of femtosecond spectroscopy for strongly correlated materials, and this reservation is naturally more apparent for even shorter duration pulses. The reason for this is simple, that the typical energy scales in quantum materials are generally much smaller than those inside molecules or atomic species, and thus the dynamics are commensurately slower. Many quantum materials can be understood largely considering only the electronic state in a narrow, few meV region around the Fermi level (Figure 2); examples include both conventional and high-temperature superconductors (59,60), where Cooper pairs are formed by coupling of electrons with equal but opposite momenta near the Fermi level (in conventional superconductors, within one phonon energy quanta) (60,61), the Dirac cone band structure in graphene and other 2D materials (62), the flat-bands in Moiré materials (63), and the quasi-particle behaviour of Fermi liquids (64). Other states where coupling to phonons or magnons is critical, e.g. charge-density waves (65,66), typically show even slower dynamics (3,36). While in all these cases other electronic effects may manifest when considering more widely spaced bands, these effects are not necessarily consequential for the interesting material properties under many circumstances. The obvious exception is under the effect of photoexcitation, where electrons can be promoted out of lower-lying bands and into higher unpopulated ones (67,68), but this is often tangential to the emergent physics. What then is the role of extremely short duration pulses in studying quantum materials? The question is timely as though attosecond pulses were first measured in 2001, attosecond-pump attosecond-probe measurements in solids have yet to be demonstrated to date, with bright enough attosecond pulses to act as pumps emerging only recently (69).

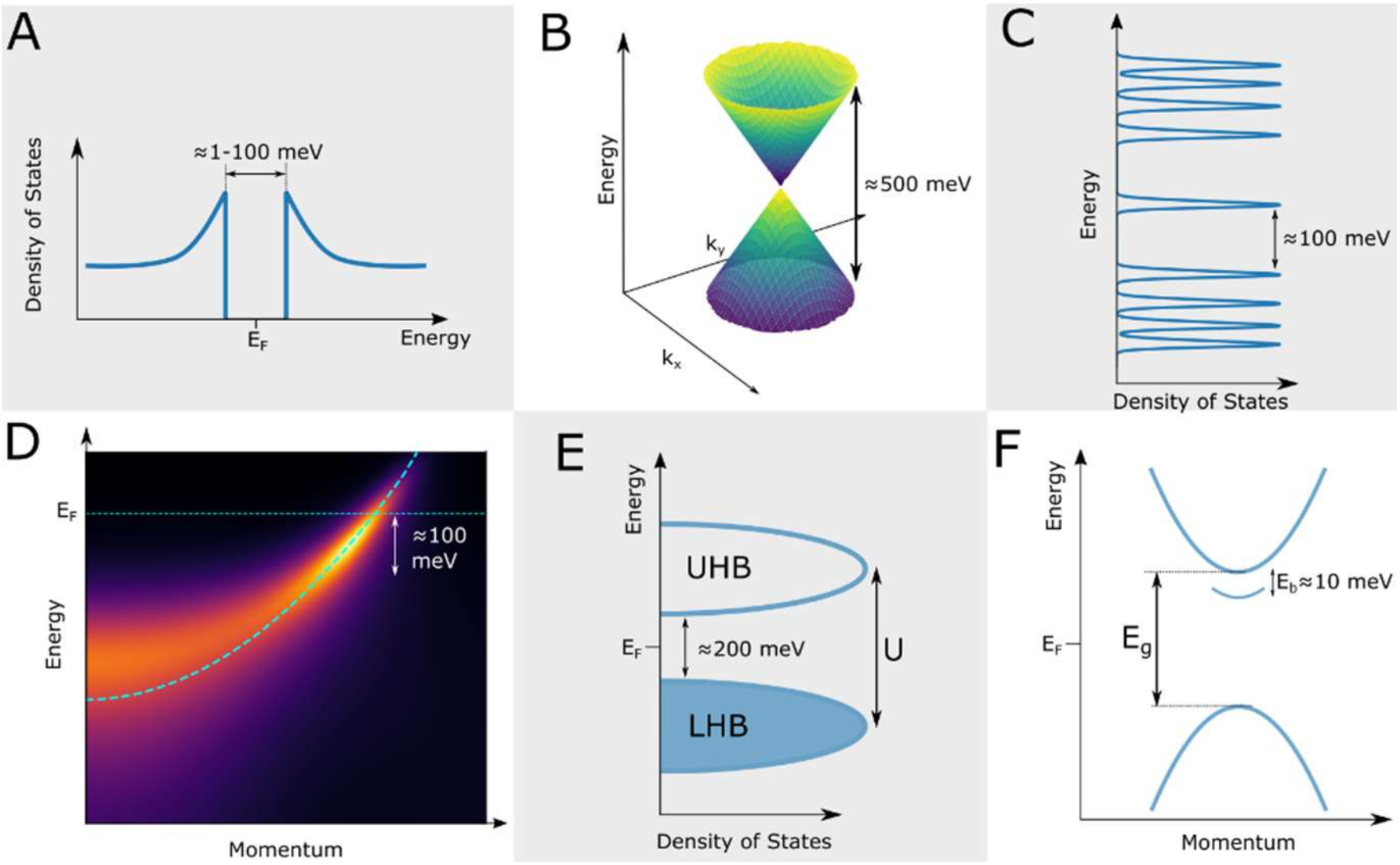

***Figure 2: Electronic energy scales in quantum materials**. (A) The superconducting gap ranges from meV level in conventional BCS superconductors to around 100 meV in the highest Tc cuprates. (B) The Dirac cone in graphene is gapless at the Dirac point, but maintains it's linear dispersion relationship for approximately 500 meV around the Fermi energy. (C) Landau levels in the presence of large multi-Tesla magnetic fields can reach splittings of 100s of meV, leading to fractional quantum Hall effects and other exotic physics. (D) In Fermi liquids the electron-like quasiparticles remain well defined within hundreds of meV of the Fermi level, before the particle-like approximation breaks down. (E) In Mott-Hubbard insulators the gap between the upper Hubbard band (UHB) and lower Hubbard band (LHB) can reach hundreds of meV, determined by the strength of the electron correlations U. (F) Excitons inside quantum materials can exhibit binding energy of tens to hundreds of meV.*

There are three cases where I foresee ~1 fs duration pulses or below having a significant impact, as illustrated in Figure 3: 1. in studying phase transitions and nonlinear dynamics, where the system can be accelerated to exhibit fast dynamics in normally slow degrees of freedom; 2. in out-running damage effects, analogous to the diffract-before-destroy approach used in X-ray crystallography, but here applied to delicate quantum states and

3. directly clocking electron-electron correlations, in which the relevant normal state dynamics may be slow but the fast response following photoexcitation can nevertheless reveal how energy moves through the system.

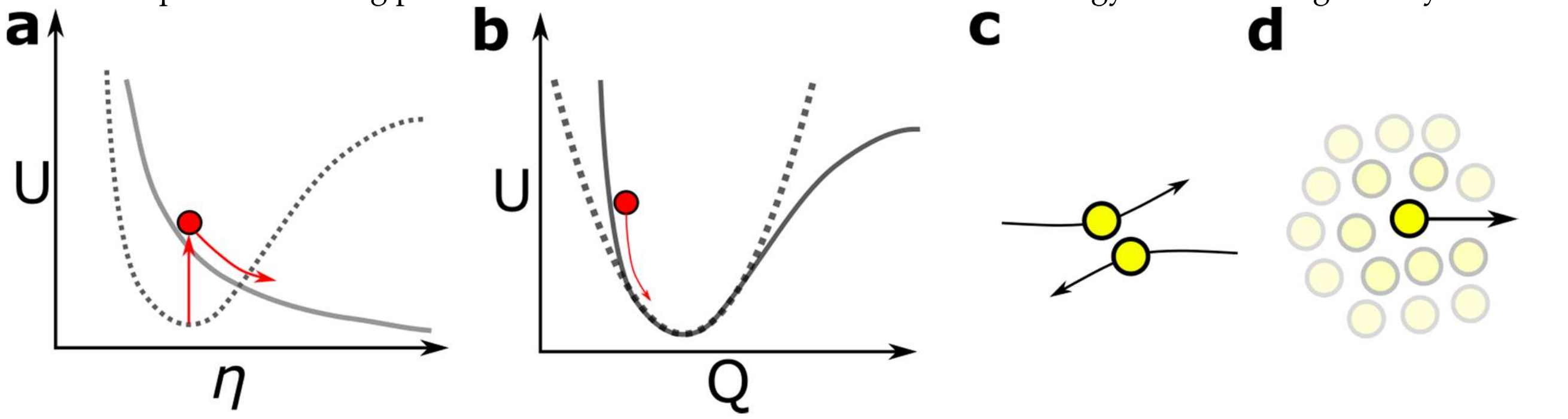

*Figure 3: Examples of fast dynamics in quantum materials. (a) During a phase transition a materials order parameter potential (dashed line) can rapidly shift, leaving the material at its original equilibrium coordinate and far from the new equilibrium position in the excited potential (dark line) that can rapidly accelerate it. (b) Lattice coordinates directly driven far from their equilibrium position, as in nonlinear phononics experiments, can sample steeper parts of the full potential (dark line) than the conventional harmonic approximation (dashed line) and accelerate to velocities higher than the normal phonon modes. (c) Electron-electron scattering and screening (d) both occur on the attosecond to few femtosecond timescale mediated by the Coulomb interaction between the light particles.*

Let us first consider the fast dynamics that can occur during phase transitions. Phase transitions are a highly general concept, with multiple different classification schemes in effect. In general they involve a discontinuous change in some generalized parameterization of the system known as an order parameter for a 1st order transition, or in it's derivative for a 2nd order transition. This order parameter could represent various types of symmetries (70). In quantum materials ultrafast photoexcitation has been used to trigger melting of charge-density wave orderings (10,15), insulator-to-metal transitions (71–73) , paramagnetic to ferromagnetic transitions, and many more (3,37). These symmetry changes are associated with significant modifications of the underlying free energy landscape (36), for instance changes from double well to single well potentials for structural dimer breaking (74,75). For 1st order phase transitions (discontinuous in the order parameter equilibrium coordinate) these changes can mean the material can be greatly displaced by a rapid excitation, and rapid nonlinear motion ensues (Figure 3a). As an example we have recently shown that photoexcited vanadium dioxide undergoes a complete reconfiguration of the electronic structure, going from insulator to metal, in less than 10 fs (29). This then launches highly nonlinear lattice motion on a timescale much faster than the relevant phonon modes, completing the transition in just a few tens of femtoseconds (76). This effect could potentially be quite general; in $VO_2$ the dimers of the vanadium ions displace by only 0.3 angstroms when changing from monoclinic to rutile structure (77), it does need seem to show exceptional large phonon acceleration or velocities within the normal phase, and larger displacements are seen in many other structural phase transitions. Notably in $VO_2$ it was found this highly out-of-equilibrium passage lead to the transient emergence of semi-metallicity (29), a state normally never found in the equilibrium phase diagram (78). Such emergent effects might well be found in a wide range of systems on very short timescales. It would also be interesting to see if ultrafast nonlinear motion could be found in other degrees of freedom. For instance, it would be exceptionally interesting in coupled electronic-magnetic transitions, as magnetism is normally considered as a relatively slow and linear process where displacements resulting in coherent precession or oscillations (36). In general the large discontinuous evolutions associated with ultrafast phase transitions offer an opportunity to apply extreme forces to different degrees of freedom inside quantum materials, driving the need for high-temporal resolution spectroscopies which have thus far only been applied to a handful of phase transitions. Ferroelectric systems like $BaTiO_3$ could be a promising target system, as ferroelectric distortions often result in large displacements within the unit cell (26,79).

Conceptually similarly, high amplitude driving of resonant modes can in turn drive nonlinear motion at speeds far higher than would be expected from examining the low-amplitude equilibrium excitations of the system (Figure 3b). This has been clearly demonstrated in the field of ultrafast phononics, where high intensity infrared radiation resonant to specific phonons can drive the lattice coordinate far away from the equilibrium position (80). Away from this position, where the motion is well described by a quadratic potential with linear phonon excitations, the potential becomes highly nonlinear (81) and the nuclei may move at much higher speeds and exhibit strong modulations of coupled degrees of freedom (80), though these speeds are still generally slow compared to the few-femtosecond timescale. Higher frequency phonons, like Jahn-Teller distortions on the other hand, can exhibit periodicities of tens of femtoseconds even in the linear regime (68,82), and would benefit from a few-femtosecond resolution (75,83,84). Jahn-Teller distortions are critical to understanding the physics of many types of perovskite complexes (85), where the extension of the oxygen cage leads to the energy separation of the different orbitals through the crystal field, which depending on the filling can lead to a metal to insulator phase transition (86,87). Because these modes are Raman active they cannot be driven directly by light to high amplitudes via a dipole coupling to trigger non-linear motion, instead being mediated by either electronic excitations or IR active phonons, but it may be possible in the future via these nonlinear phonon couplings.

Perhaps the most natural application for attosecond pulses is in examining electron correlations (88). While as previously pointed out the energy scales for electrons in quantum materials can often be quite small, leading to slow dynamics, following photoexcitation electrons can be promoted to higher energy states and evolve quickly through interactions with other electrons. For instance attosecond photoemission experiments have been used to distinguish between electron-electron scattering and screening effects (Figure 3c and d) via the attosecond lifetimes of intermediate excited states (57). Understanding electron scattering processes is the heart of the so-called strange metallic phase (89) observed in superconducting cuprates and other materials. In strange metals the resistivity scales linearly with temperature, as opposed to quadratically for normal metals, and is associated with a linear scaling of the charge scattering time over an extremely, broad temperature range (90). Indeed strange metals, like bad metals, break the so-called Mott-Ioffe–Regel limit where the mean free path reaches the single unit cell level (90,91), implying a breakdown in conventional charge carriers and charge scattering. Intriguingly the scattering time in many cuprates has been shown to correspond to maximum entropy dissipation or Planckian dissipation, $\tau = \hbar/_{k_B T}$ (92). This implies scattering times of just 25 fs at room temperature, necessitating the use of ultrashort pulses to study dissipation in the time domain. Strange metals are known to exhibit several other intriguing properties like strongly suppressed electronic shot noise (93), an absence of quasiparticle excitations (90), and atypical optical conductivities (94) which further motivate time-resolved of the internal electronic correlations. Because anomalous scattering times in the optical conductivity have been related back directly to the same underlying quantum critical behaviours as leading to the linear conductivity and dissipation time (94), time-domain probing of these processes could reveal the scattering mechanisms responsible. Previous ultrafast studies on cuprates have revealed interesting correlations but have focused on bosonic couplings rather than strange metallicity (95–97). Strong electron correlations that could lead to fast electron-electron scattering are also expected in more generally in Mott insulators as well as flat band systems.

There are several barriers to the use of attosecond and few-cycle sources to the study of strong electron correlations inside quantum materials, however. When electrons are excited by short pulses they are simultaneously promoted from many different energy levels across the valence bands to many different states in the conduction band or continuum, with decreasing time resolution scaling inversely to the increasing bandwidth, leading to a complex evolving distribution which could involve numerous simultaneous correlation functions. This will tend to wash out any dynamics and lead to simple exponential decay in the macroscopic signal, and prevent reconstruction of the underlying microscopic dynamics from experimental observables. One could envision using high-temporal resolution multi-dimensional spectroscopy (98,99) to separate out multiple channels, but even in this case the multiple potential degeneracies (from spectrally overlapping signal channels) would make interpreting the signals challenging. This closely relates to another challenge, the need for improved theoretical models. The early experiments alluded to earlier distinguishing scattering from screening (100) and other experiments looking at attosecond melting of a Mott insulator (101), together with other attosecond studies of condensed phase systems (52,102), required extensive modelling to interpret the signals, and at present few frameworks exist to explain ultrafast probes of multi-electron dynamics other than direct intensive simulations. This is naturally a challenge for quantum materials, where even simulating the optical properties may be challenging (103). Though many successful methods for simulating material dynamics like nonequilibrium Green-function methods, nonequilibrium dynamical mean-field theory, time dependent density-functional theory(37), and tensor-network methods now exist (37,76), the development of robust generic experimental and theoretical frameworks for studying electron correlations – agnostic to the details of the particular material under study - is a clear frontier for ultrafast studies in quantum materials. Recent descriptions of new types of multidimensional spectroscopies move in this direction (104,105), as it is clear the complexity inherent to quantum materials cannot reasonably be mapped to low-dimensional information like emission spectra directly.

Finally, related to the use of attosecond methods is the advent of lightwave control and lightwave spectroscopy experiments, where a strong driving field is used and the system under study responds not to the overall pulse envelope but to the oscillations of the electric field (106). The bulk of attosecond measurements in solids thusfar have taken this approach, whether in pump-probe experiments (7,52,102,107,108) or via high-harmonic generation spectroscopy (51,109,110), in part due to the difficulty in providing sufficiently short duration pump-pulses to enable conventional pump-probe schemes at sub-femtosecond timescales. Like in the gas phase (111,112), the challenge here is that the strong driving field can substantially modify the material under study, and so the system under study is often better thought of as hybrid light-matter system, as in the emerging field of Floquet quantum materials (23,37,113,114). This is a fascinating topic when viewed through the lens of quantum control and engineering, but layers on additional complexities when used to understand complex systems that, in my opinion, limit it as a widespread approach for quantum materials. I will however briefly note that lightwave approaches are, through their highly nonlinear nature, exquisite probes of material symmetry through their polarization dependence (115), and could be widely applied for this application without requiring such intensive modelling.

An entirely different regime where attosecond pulses may be of great utility is in mitigating material damage. Radiation damage, including heating, is an inevitable consequence of photoexcitation, whether extremely gentle, as in the case of THz radiation, or highly affective, as with X-rays. In quantum materials the specific phases or orderings may be extremely delicate, with many emerging only at mK temperatures (116,117). In these cases any kind of probe, whether optical or otherwise, can reasonably be expected to strongly disturb or even destroy the ordering (118), whether by heating or other changes following photoabsorption, depending ultimately upon the dose. This is very similar to the challenge faced in X-ray diffraction measurements of proteins, in which radiation damage gradually degrades the crystal structure (119). As both damage and signal are directly linked to exposure, this was long thought an intractable problem before it was realized that the use of a very short and bright X-ray pulse like the type produced at X-ray free electron lasers could allow for a diffraction pattern to be acquired from a single high brightness pulse (120). Following this intense pulse the sample would inevitably be destroyed, but provided the pulse was significantly shorter than this cascaded destruction process high-quality images could be produced (121). It has been proposed that the same concepts can be applied not to structural orders but also to electronic ones by using recently developed attosecond X-ray pulses, including delicate quantum states (122).

In a recent landmark study at SwissFEL, Vonka *et al.*, have shown the same concepts apply to condensed phase systems, demonstrating that ~5 fs gating of X-ray diffraction is required for measurements of the intrinsic valence electronic structures in silicon (123), a process which they assign to electronic dephasing prior to lattice motion. Longer exposures to moderate intensity X-rays – but still well below the damage threshold - prior to scattering lead to distorted structure factors, and could thus impact the interpretation of ultrafast resonant diffraction experiments that have been used to study magnetic orderings (44,124), charge and orbital ordering (125), and charge density wave orders (126), among many others. The effect may reasonably be expected to be more significant for delicate emergent orders at lower temperatures, and so the application of attosecond X-ray pulses will likely become an essential tool for the quantitative understanding of emergent electronic order in quantum materials.

**High-Repetition Rate Sources**

High-repetition rate sources, moving beyond typical kHz repetition rates towards 100s of kHz, MHz, or even GHz repetition rates provide a simple key advantage – higher fluxes while maintaining peak powers below those distorting the measurement, which may be distorted for a number of reasons discussed later. This allows weak signals to be extracted in a somewhat brute force way where applying 10x more pulses leads to getting 10x more signal and a $\sqrt{10}$ increase in signal-to-noise. A case where this is an especially good approach, as opposed to just increasing energy per pulse, is for photoelectron measurements.

In photoelectron method using low pulse energies is essential. This is because of space-charge effects; when two electrons are emitted there is a mutual Coulomb repulsion between them, which increases as they are emitted closer together in space or time (127). For all but the simplest photoelectron methods it is important to accurately recover the energy or momenta of the emitted electrons, and space-charge degrades this significantly. For instance in angle resolved photoemission spectroscopy (ARPES) the energy and momentum of photoemitted electrons from a surface are measured to provide a reconstruction of the materials band structure. Mutual repulsion between emitted electrons modifies these parameters and gives rise to distorted band structures (128), The problem is even more pronounced for spatially-resolved nano-ARPES measurements where photoemission is restricted to a small region via tight focusing of the ionizing radiation, increasing proximity effects (129). To completely remove space-charge effects it is thus desirable to restrict photoemission to one electron per pulse; accounting for shot noise, this means a mean electron yield per pulse significantly below one. ARPES and other photoelectron spectroscopies are thus ideal candidates for high-repetition rate HHG sources in particular, where the relatively low-yield per pulse is actually a great strength. High-repetition rate HHG sources for ARPES are now widespread, but significant tradeoffs remain in terms of achievable photon energy, stability is a significant issue for time-resolved studies, and MHz sources are only now emerging (130,131).

What are the limits to higher-repetition rates for photoelectron spectroscopies? The foremost, which is the case for all high-repetition rate applications, is heating and damage. At synchrotron sources, where pulses tend to arrive at MHz repetition rates, this is already a commonly faced problem. Heating can be partially curbed by studying epitaxially grown films on highly thermal conductive substrates (132), but this naturally limits the range of systems that can be studied, especially as nano-ARPES moves into studying quantum materials in devices (133). Active cooling can also assist, decreasing the base temperature to counteract the effects of the photothermal load. Another problem, perhaps less obvious than heating, is charge replenishment. It is already well known that PES is not well suited to insulating materials because as electrons are extracted from the surface, new ones cannot move through the material to replenish them, and even in semiconducting materials this can lead to distortion (134). At higher repetition rates this could become a problem in a much wider range of systems. In metals electrons can move at significant fractions of the speed of light, with Fermi velocities of around 0.01c (135), so a ten micron region can replenish its emitted electrons in just 10s of picoseconds under ideal conditions. Conversely in semiconductors or diffusive metals, which tend to describe most strongly-correlated systems, the electron replenishments speeds can be much lower than the Fermi velocity. In proposed slow metals it could be smaller than the speed of sound (136), leading to replenishment times of several nanoseconds. While electrons can also be replenished through beam injection or other methods (137) these are unlikely to be practical, and for many systems MHz repetition rates appear to be a reasonable limit.

Of course higher repetition rates are generally desirable for all systems. The general limitations are related to sample recovery. For instance, photoinduced phase transitions commonly have recovery times of microseconds or even milliseconds (132,138), and applying lower fluence excitation is not possible while still driving the transition. In this case it is possible to scan the sample and excited fresh points with each shot, a proven approach at low repetition rates (138,139). As the repetition rate increases, so does the required scanning speed. For instance, probing a fresh 10 micrometer spot at MHz repetition rate requires 10 m/s scanning rates, challenging for linear stages but achievable with a sample mounted on a kHz repetition rate rotation stage (138). This of course in turn poises significant challenges to sample preparation and experimental apparatuses, like whether sufficiently large and homogenous samples can be prepared, and whether the experimental system can actually interface with such a moving stage – particularly a problem for cryogenic samples. Nevertheless, the hard limits for sample scanning appear to be in the 10s to 100s of MHz range, and so there is still significant room for improvement in terms of repetition rates.

**High-Field THz and IR Sources**

THz and mid-to-far IR radiation is particularly interesting for driving quantum materials as it can provide below bandgap driving, minimizing photoexcitation, while often being resonant to important vibrational or emergent collective modes (37,39,53). As such high-field THz/IR radiation has the potential to drive interesting non-linear phenomena in quantum materials while remaining below the damage threshold. This brings us very close to using light to control materials, but I will endeavour to remain within the realm of spectroscopy, simply to keep the scope of this perspective reasonable. It is worth noting that long-wavelength radiation *has* been used to drive materials above the damage threshold in the past, notably using bowtie nano-antennas to boost local THz field strength (140), but is usually non-damaging in semi-conducting or insulating materials even up to ~100 MV/cm field strengths in the THz (141). As such there is still considerable scope for moving to higher peak fields in the THz, especially in concert with tuneable frequency, as some of most widespread high-peak field generation methods in the THz generally operate at fixed frequencies (142). This is certainly not universally true, as plasma and free-electron laser sources offer great tune-ability, but these come with their own limitations related to scale.

In the mid to far IR there are also gaps where efficient sources are lacking. Here the use case is perhaps more readily apparent, to address specific phonon modes for nonlinear phononics (26,80,143) and driving phase transitions (39,144). Conversely, what opportunities would be opened by higher field, tuneable THz sources? A large number of examples of exotic quantum materials phenomena revealed by THz radiation exist in the literature, for example exposing optical silent Josephson resonances between superconducting stripes in the cuprate superconductors (145,146) or Landau levels in topological systems (35). A very elegant example is non-linear Higgs spectroscopy (5,147), where THz radiation drives the collective Higgs excitation of Cooper pairs in a high-temperature superconductor. This is a third-order process in the THz field which is generally quite weak, but similar spectroscopies could be applied to other collective modes, for instance magnon condensates.

Perhaps most exciting of all is how high-field THz sources now enable 2D spectroscopy of low-energy excitations, separating out different contributions to scattering and dephasing in optical signals (148,149). For example this approach was recently applied studying superconducting gap dynamics (148,150), to separate out dephasing and loss processes in the electronic degree of freedom in a strongly correlated electronic systems (151), and to strange metal phases, where both quasi-particle-like and Planckian relaxation processes were observed to contribute to the overall optical response in parallel (152,153). In systems exhibiting strong inhomogeneities and sensitivities to perturbations like quantum materials this multidimensional approach, capable of removing the impacts of inhomogeneous linewidth broadenings and separating out different scattering processes, may be exceptional valuable moving forward. These are exceptionally photon hungry processes, and so the development of higher brightness and field strength sources will continue to drive the field forward in the years to come.

**Quantum Light**

Quantum light – photons exhibiting non-classical states in photon number or other degrees of freedom – have been available for many years but have largely failed to make a significant impact on condensed matter physics thusfar. However, the development of ever more robust generation schemes and the potential for non-classical states with much higher photon numbers (154) than those traditionally generated by spontaneous parametric down conversion, means that significantly more attention is being drawn in this direction.

The practical use of quantum light for metrology was first demonstrated by the use of squeezed light to reduce shot noise in gravitational wave detectors, where extremely high precision in interferometry naturally benefited from phase-squeezing (155). Such a reduction in shot noise is generally useful, but is unlikely to be transformative for quantum materials studied with optical fields. As quantum materials are usually robust to optical excitation, it is typically simpler to acquire more signal and reduce the relative noise through the $\sqrt{n}/n$ scaling, where n is the number of photons detected. Improving noise with photon statistics could be significantly more important for studies using X-rays however, where the photon numbers used are generally much smaller because of considerations like sample damage, detector saturation and intrinsically low cross sections. For instance resonant inelastic X-ray scattering processes (RIXS) often have cross sections of order $10^{-24}$ $cm^2$ (156) and acquisitions of several days may be required at the brightest synchrotron sources to beat the shot noise limit. Squeezed X-rays generated through non-linear optical process like HHG (157–159) or at XFELs (160), only recently demonstrated and not yet at a practical level, may offer a significant advantage for such problems, particularly for photon sensitive materials where increasing the total photon dose is not feasible.

Beyond photon statistics, it has been shown that non-classical photon states can enhance nonlinear optical processes. For instance with entangled photon pairs the yield of two-photon absorption and second harmonic generation can scale linearly with the optical intensity, rather than the typical quadratic (161–163). Intuitively this can be understood as the correlations ensuring both photons arrive simultaneously both temporally and spatially to interact with the system, instead of statistically. While this can be used to reduce dose, a more exciting possibility is to use them to study correlated states. An example would be Cooper pairs inside superconductors. It was previously suggested that single-photon initiated, double-electron emission could be used to study Cooper pair formation inside superconductors as the correlations between both electrons would enhance the probability of double-electron emission (164,165), with extensions to the general correlated electron problem and exotic pairings subsequently proposed (166–169). This is a rather important problem due to longstanding debate over the nature of high-temperature superconductivity. While in conventional superconductors Cooper pairs form and condense into a macroscopic wavefunction with phase coherence, enabling superconductivity, it has been suggested that in unconventional superconductors Cooper pairs form at higher temperatures but do not condense into a macroscopic wavefunction until reaching $T_c$ (170,171). The observation of Cooper pairs above Tc, with a method which does not rely on the macroscopic superconducting response, would be a significant contribution to this debate. Unfortunately the cross-section for single-photon double-emission has thusfar proven to be prohibitively small even for conventional superconductors (172,173).

More recently theory work has extended the study of double-electron emission from the single-photon to the double photon case. Key here to obtaining genuine correlation functions is ensuring that the electrons are emitted coincidentally in both space and time; this has been done by considering double-ionization from a short pulse which confines the two photoionization events in time (174,175) or via detector gating in time (176,177), and by numerically enforcing that the electrons come from the same spatial region. Practically this would enforce using small samples or illuminating regions below the correlation length in size, a daunting challenge.

Interestingly it appears that the use of entangled pairs, which naturally come with spatial and temporal correlations (162), has not yet been considered in the literature, and so I propose that using quantum light to drive two-photon two-electron emission could help overcome the challenges related to double emission cross-sections, namely accidental coincidences from independent photoemission events and final state e-e scattering. These challenges apply to both classical and quantum light driven photoemission, but the relative boost in true coincidences from quantum light could prove decisive. Using correlated light to probe correlated materials is still an emerging topic, but seems to be a promising direction to help overcome the challenges inherent to using coherent light.

Finally I note that entangled light can enable measurements that provide simultaneous high-resolution information in two complementary bases beyond the limit expected from the classical Fourier bandwidth via their joint bandwidth (178–180). By pumping spontaneous parametric down conversion using a narrowband source, broadband ultrafast signal and idler pulses can be emitted and used for ultrafast experiments. Despite these pulses being broadband, the joint spectral correlations remain narrowband, allowing coincidence measurements to provide simultaneous high temporal and spectral resolution. This ability to provide, for example, high resolution temporal measurements of electronic phase transitions while maintaining the ability to resolve the distinct bands could resolve one of the major challenges of time-resolved ARPES and ultrafast coherent imaging. Numerous other potential applications of quantum light to spectroscopy have also been highlighted, though so far primarily to molecular or nanostructured system, with comprehensive reviews available (160,179).

**Conclusions**

In this perspective I have tried to provide an overview of where next generation ultrafast sources can provide an impact on the study of quantum materials, considered as much as possible from the needs of condensed matter as opposed to the development of new methodologies or opening of new problems. For this reason I have tried to highlight specific important problems in quantum materials – strange metals, superconducting fluctuations above $T_c$, measurement of photosensitive charge-density waves, etc. – wherever possible. These systems stand to benefit from improved spectro-temporal resolution, higher-repetition rates and quantum statistics, and attosecond pulse durations, respectively, but many frontiers of ultrafast pulse generation can be associated to clear and pressing problems in quantum materials. However, it is well worth noting that the development of new spectroscopic techniques and sources can have a dramatic impact on our understanding of materials previously considered “conventional”, and redefine the research landscape. A striking illustration of this is the recent recognition of altermagnetism (181). Many altermagnetic candidates had been studied for decades and were classified as conventional antiferromagnets, and their fundamentally distinct character went unnoticed until the right theoretical framework and experimental signatures were identified (182). The lesson extends well beyond magnetism: even well-characterized materials may hide qualitatively different physics that existing probes were not designed to detect, which is a strong motivation for developing new spectroscopic tools. Furthermore, where a material is considered “interesting” is often simply a matter of how well known it becomes. For instance, while ARPES is quite limited in the range of materials it can address, the information it provides is so valuable that research effort has quite naturally become especially focused on those materials that it can address. Perhaps ultrafast spectroscopy can also hope to cause significant re-evaluation of existing material classes through the discovery of unexpected dynamical processes, for example those observed in wide bandgap systems (52,108).

The new ultrafast sources appearing now, with higher repetition rates, attosecond durations, high-field strengths at longer wavelengths, and quantum properties, will no doubt transform our understanding of strongly correlated systems in the years to come in ways difficult to predict. Of all of these, how the use of quantum light will affect our understanding is perhaps the least clear, but the most exciting. Modelling of quantum materials is increasingly carrier out using ideas from quantum information science like entanglement entropy that are indirectly connected to experimental observables; could the use of quantum light allow access to such presently abstract measures? The limit will soon no longer be our technology, but rather our imaginations. Furthermore, experiments on solids remain almost universally interpreted within the scope of band structures and field theories, models of materials which are fundamentally static in origin. As we approach studying materials at the fastest timescales and smallest lengthscales, it will be interesting to see if this framework continues to hold.

## Acknowledgments

ASJ acknowledges the support of the Ramón y Cajal Program (Grant RYC2021-032392-I), the Spanish AIE projects PID2022-137817NA-I00 and CNS2025-165235, and the Comunidad de Madrid (project TEC-2024/TEC-380 "Mag4TIC"). Funded by the European Union (ERC, KnotSeen, 101163311). Views and opinions expressed are however those of the author(s) only and do not necessarily reflect those of the European Union or the European Research Council. Neither the European Union nor the granting authority can be held responsible for them. IMDEA Nanociencia has been accredited as Centro de Excelencia Severo Ochoa, a distinction awarded by the Spanish State Research Agency for the period 2026–2031. This publication is part of the grant «Excelencia Severo Ochoa» CEX2025-001660-S, funded by MICIU/AEI/10.13039/501100011033

# Additional Information

**Data Accessibility**
This article has no data.

**Authors' Contributions**
Allan S. Johnson is wholly responsible for this perspective.

**Competing Interests**
*I have no competing interests.*